\documentclass[draft]{agujournal}
\draftfalse

\journalname{Geophysical Research Letters}

\graphicspath{{./}{./figs/}}

\begin{document}

%
%


\title{Abundances of uranium and thorium elements in Earth estimated by geoneutrino spectroscopy
}

%
%




\authors{
S.~Abe\affil{1},
S.~Asami\affil{1},
M.~Eizuka\affil{1},
S.~Futagi\affil{1},
A.~Gando\affil{1},
Y.~Gando\affil{1}\thanks{Department of Human Science, Obihiro University of Agriculture and Veterinary Medicine, Obihiro, Hokkaido 080-8555, Japan},
T.~Gima\affil{1},
A.~Goto\affil{1},
T.~Hachiya\affil{1},
K.~Hata\affil{1},
K.~Hosokawa\affil{1}\thanks{Kamioka Observatory, Institute for Cosmic Ray Research, The University of Tokyo, Higashi-Mozumi, Kamioka, Hida, Gifu 506-1205, Japan},
K.~Ichimura\affil{1},
S.~Ieki\affil{1},
H.~Ikeda\affil{1},
K.~Inoue\affil{1},
K.~Ishidoshiro\affil{1}, 
Y.~Kamei\affil{1},
N.~Kawada\affil{1},
Y.~Kishimoto\affil{1,2}, 
M.~Koga\affil{1,2}, 
M.~Kurasawa\affil{1},
N.~Maemura\affil{1}, 
T.~Mitsui\affil{1},
H.~Miyake\affil{1},
T.~Nakahata\affil{1}, 
K.~Nakamura\affil{1}, 
K.~Nakamura\affil{1,2}\thanks{Faculty of Health Sciences, Butsuryo College of Osaka, Sakai, Osaka, 593-8328, Japan},
R.~Nakamura\affil{1},
H.~Ozaki\affil{1,3},
T.~Sakai\affil{1},
H.~Sambonsugi\affil{1},
I.~Shimizu\affil{1}, 
Y.~Shirahata\affil{1}, 
J.~Shirai\affil{1}, 
K.~Shiraishi\affil{1},
A.~Suzuki\affil{1},
Y.~Suzuki\affil{1},
A.~Takeuchi\affil{1},
K.~Tamae\affil{1}, 
H.~Watanabe\affil{1}, 
Y.~Yoshida\affil{1},
S.~Obara\affil{4}\thanks{National Institutes for Quantum Science and Technology (QST), Sendai 980-8579, Japan},
A.~K.~Ichikawa\affil{5}, 
S.~Yoshida\affil{6},
S.~Umehara\affil{7},
K.~Fushimi\affil{8}, 
K.~Kotera\affil{8}, 
Y.~Urano\affil{8},
B.~E.~Berger\affil{9,2},
B.~K.~Fujikawa\affil{9,2},
J.~G.~Learned\affil{10},
J.~Maricic\affil{10},
S.~N.~Axani\affil{11},
Z.~Fu\affil{11},
J.~Smolsky\affil{11},
L.~A.~Winslow\affil{11},
Y.~Efremenko\affil{12,2},
H.~J.~Karwowski\affil{13,14},
D.~M.~Markoff\affil{13,15},
W.~Tornow\affil{13,16,2},
A.~Li\affil{14},
J.~A.~Detwiler\affil{17,2},
S.~Enomoto\affil{17,2},
M.~P.~Decowski\affil{18,2},
C.~Grant\affil{19},
H.~Song\affil{19},
T.~O'Donnell\affil{20},
S.~Dell'Oro\affil{20}
\\(The KamLAND Collaboration)}



\affiliation{1}{Research Center for Neutrino
    Science, Tohoku University, Sendai 980-8578, Japan}
\affiliation{2}{Institute for the Physics and Mathematics of the Universe, The University of Tokyo, Kashiwa 277-8568, Japan}
\affiliation{3}{Graduate Program on Physics for the Universe, Tohoku University, Sendai 980-8578, Japan}
\affiliation{4}{Frontier Research Institute for Interdisciplinary Sciences, Tohoku University, Sendai, 980-8578, Japan}
\affiliation{5}{Department of Physics, Tohoku University, Sendai 980-8578, Japan}
\affiliation{6}{Graduate School of Science, Osaka University, Toyonaka, Osaka 560-0043, Japan}
\affiliation{7}{Research Center for Nuclear Physics (RCNP), Osaka University, Ibaraki, Osaka 567-0047, Japan}
\affiliation{8}{Graduate School of Advanced Technology and Science, Tokushima University, Tokushima, 770-8506, Japan}
\affiliation{9}{Nuclear Science Division, Lawrence Berkeley National Laboratory, Berkeley, CA 94720, USA}
\affiliation{10}{Department of Physics and Astronomy, University of Hawaii at Manoa, Honolulu, HI 96822, USA}
\affiliation{11}{Massachusetts Institute of Technology, Cambridge, MA 02139, USA}
\affiliation{12}{Department of Physics and Astronomy, University of Tennessee, Knoxville, TN 37996, USA}
\affiliation{13}{Triangle Universities Nuclear Laboratory, Durham, NC 27708, USA}
\affiliation{14}{The University of North Carolina at Chapel Hill, Chapel Hill, NC 27599, USA}
\affiliation{15}{North Carolina Central University, Durham, NC 27701, USA}
\affiliation{16}{Physics Department at Duke University, Durham, NC 27705, USA}
\affiliation{17}{Center for Experimental Nuclear Physics and Astrophysics, University of Washington, Seattle, WA 98195, USA}
\affiliation{18}{Nikhef and the University of Amsterdam, Science Park, Amsterdam, The Netherlands}
\affiliation{19}{Boston University, Boston, MA 02215, USA}
\affiliation{20}{Center for Neutrino Physics, Virginia Polytechnic Institute and State University, Blacksburg, VA 24061, USA}





\correspondingauthor{Nanami Kawada}{kawada@awa.tohoku.ac.jp}




\begin{keypoints}
\item Geoneutrino measurement with low reactor neutrino backgrounds improves the distinct spectroscopic contributions of U and Th
\item Radiogenic power in the Earth estimated from this geoneutrino measurement is consistent with a range of models and disfavors the higher power model
\item Identifying the Earth's mantle contribution to the total geoneutrino flux strongly depends on an accurate estimation of the crustal contribution
\end{keypoints}

%
%


\begin{abstract}
The decay of the primordial isotopes $^{238}\mathrm{U}$, $^{235}\mathrm{U}$, $^{232}\mathrm{Th}$, and $^{40}\mathrm{K}$ have contributed to the terrestrial heat budget throughout the Earth's history.
Hence the individual abundance of those isotopes are key parameters in reconstructing contemporary Earth models.
The geoneutrinos produced by the radioactive decays of uranium and thorium have been observed with the Kamioka Liquid-Scintillator Antineutrino Detector (KamLAND).
Those measurements have been improved with more than an 18-year observation time, and improvement in detector background levels mainly with an 8-year nearly reactor-free period, which now permit spectroscopy with geoneutrinos.
Our results yield the first constraint on both uranium and thorium heat contributions.
The KamLAND result is consistent with geochemical estimations based on elemental abundances of chondritic meteorites and mantle peridotites.
The High-Q model is disfavored at 99.76\% C.L. and a fully radiogenic model is excluded at 5.2$\sigma$ assuming a homogeneous heat producing element distribution in the mantle.
\end{abstract}

\section*{Plain Language Summary}
The energy to drive the Earth's engine comes from two different sources: primordial and radiogenic.
Primordial energy comes from the added heat by collisions of accreting material and less so by the energy accompany the sinking of metal to form the core.
The radioactive decays of heat producing elements ($\it{i.e.}$ potassium, thorium, and uranium) also generate energy and some of these decaying elements produce antineutrinos (geoneutrinos).
Geoneutrino measurements provide the Earth's fuel gauge for its radiogenic power supply and insights into the planet's cooling history.
The measurement accuracy of the KamLAND experiment has been improved by an 18-year long-term observation and a reduction of the significant background generated by commercial reactors.
Consequently, modern geoneutrino measurements have entered an era of distinct spectroscopic contributions coming from uranium and thorium.
The KamLAND result is consistent with compositional models for the bulk silicate Earth (the crust plus the mantle) predicting low to medium radiogenic heat (10 to 20 TW ($10^{12}$ watts)) and disfavor high concentration models (30 TW).  This constraint sets the best limit on the permissible radiogenic energy budget in the Earth.
Geoneutrino observations now begin to make significant contributions to the understanding of fundamental driving forces powering the Earth dynamic behaviour.

\section{Introduction}

The heat flow inside the Earth's interior is central to our understanding of geophysical processes.
Radiogenic heat from the decay of $^{238}\mathrm{U}$, $^{235}\mathrm{U}$, $^{232}\mathrm{Th}$, and $^{40}\mathrm{K}$ provides a significant fraction of the Earth's total heat budget.
Indirect estimates of the radiogenic heat fraction from geophysics or geochemistry have large systematic uncertainties.
Instead, the measurements of electron antineutrinos ($\overline{\nu}_\mathrm{e}$) produced in beta decays in the decay process of these elements provide a powerful tool for direct investigation.
Due to the weakly-interacting nature of the neutrino, these neutrinos from radiogenic decay, called ``geoneutrinos'' represent a unique probe to directly measure the fraction of radiogenic heat in the Earth's mantle and therefore offer an insight into the convection in this region.
New insight into the mantle radiogenic heat opens up a new window into further understanding of mantle convection, plate tectonics, and the geodynamo.
  
The first step in making neutrino geophysics a reality was taken by the Kamioka Liquid-Scintillator Antineutrino Detector (KamLAND) Collaboration reporting on the initial detection of electron antineutrinos produced by decays within the Earth~\citep{Araki2005b}.
Since then, the Borexino experiment at the Laboratori Nazionali del Gran Sasso confirmed the detection of geoneutrinos~\citep{Bellini2010}, by providing a complementary geological measurement in Italy to that of KamLAND in Japan.
The measurements from these two experiments have continued to improve, and in 2011 KamLAND presented the first radiogenic heat estimation from uranium and thorium based on the observed $\overline{\nu}_{e}$ flux~\citep{Gando2011}. 

KamLAND was originally designed to measure the $\overline{\nu}_{e}$'s with energies of a few MeV from 56 commercial nuclear reactors, at a mean weighted distance of 180\,km.
The deficit in the number of measured electron antineutrinos demonstrated that reactor $\overline{\nu}_{e}$ oscillate, disappearing (becoming muon and tau neutrinos which are not detected at these energies) and reappearing over three cycles.
This measurement confirmed the evidence for the same oscillation mechanism proposed to explain the long-standing deficit in electron neutrinos from the sum (as opposed to electron antineutrinos from power reactors).
However, the same high reactor $\overline{\nu}_{e}$ flux that was necessary for the discovery of $\overline{\nu}_{e}$ oscillation, also creates the largest background and source of uncertainty for the geoneutrino measurement.
This situation changed drastically after the Fukushima-I reactor accident in March 2011, and subsequent long-term shutdown of Japanese nuclear reactors.
Despite being the result of a great tragedy, this situation provided an opportunity to measure the geoneutrino flux more precisely.
In particular, we are able to perform geoneutrino spectroscopy, separating the uranium and thorium contributions, thus probing the abundance of the major heat producing elements.
In the following sections, we outline our Earth composition model, provide the details of the analysis and discuss the implications for these results, including new constraints on the radiogenic heat fraction in the mantle.

\section{Earth Composition Model and Geoneutrino Flux Prediction}
In order to interpret the KamLAND data, a prediction of the geoneutrino flux is needed.
This in fact requires a model of the location and abundance of the isotopes of interest: $^{238}$U, $^{232}$Th, and $^{40}$K.
In general, these models use a shell structure established by analyses of seismic wave propagation, a bulk chemical composition estimated by compositional analyses of the primitive meteorites which are the remnants of the nebula that formed our solar system, and chemical analysis of rock samples.
One of the leading models is the Bulk Silicate Earth (BSE) model that provides estimates for the elemental abundances of refractory lithophile elements such as uranium and thorium, based on the studies of measured abundances of CI carbonaceous chondrites and mantle peridotites.
The estimate of volatile elements such as potassium have a relatively large uncertainty in this model since they became depleted during Earth's accretion.
The BSE model gives average composition predictions for the primary heat-producing elements $^{238}$U ($T_{1/2} = 4.47$\,Ga), $^{235}$U ($T_{1/2} = 0.71$\,Ga), $^{232}$Th ($T_{1/2} = 14.0$\,Ga), and $^{40}$K ($T_{1/2} = 1.28$\,Ga).
The BSE heat production estimates are 8\,TW from the $^{238}$U decay series, 0.3\,TW from the $^{235}$U decay series, 8\,TW from the $^{232}$Th decay series, and 4\,TW from $^{40}$K~\citep{McDonough1995,Lyubetskaya2007,Arevalo2009}.
This present-day radiogenic heat is nearly half of the Earth's total heat flow ($47 \pm 2$\,TW)~\citep{Davies2010}.

The present work is based on the KamLAND data and a simple BSE-inspired reference model~\citep{Enomoto2007} with minor updates described below to reconstruct the radiogenic element distribution. 
The crustal composition model used in~\citep{Enomoto2007} has been updated by changing the U and Th concentrations in the upper continental crust by $-3.6$\% and $-1.9$\%, respectively, and those in the middle continental crust by $-19$\% and $+6.6$\%, respectively~\citep{Rudnick2014}. The neutrino flux from those reservoirs has been scaled accordingly. 

Due to their high permeability in matter, geoneutrinos provide unique and direct information of the Earth's interior.
However, given the terrestrial propagation distance the phenomenon of $\overline{\nu}_{e}$ oscillation must be included.
Neutrino oscillation depends on the $\overline{\nu}_{e}$ energy and distance traveled.
Because of the dispersed $\overline{\nu}_{e}$ source position, we use an average ``survival'' probability $P_{ee} = 0.554^{+0.012}_{-0.009}$ obtained from the latest neutrino oscillation data~\citep{Gando2013}.

\section{Geoneutrino Detection}

KamLAND is located in the Kamioka mine under Mount Ikenoyama at a depth of $\sim$2700\,m water-equivalent.
The rock overburden effectively shields against cosmic-ray induced atmospheric muons.
The central part of the detector housed in a 18-m-diameter stainless steel spherical tank is sensitive to neutrino interactions and consists of 1\,kton of ultra-pure liquid scintillator~(LS) contained in a 13-m-diameter spherical balloon made of 135-$\mu$m-thick nylon-based transparent film, which is surrounded by transparent buffer oil (BO).
LS is basically a mixture of long chain and aromatic ring hydrocarbons with an H:C proportion of $\sim$2 where hydrogen is the target for $\overline{\nu}_{e}$ detection.
The scintillation photons are viewed through BO by an array of 1325 fast 17-inch-diameter photomultiplier tubes (PMTs), and 554 slow 20-inch PMTs bolted on the inner surface of the stainless steel tank, providing 34\% solid-angle coverage in total.
This inner detector (ID) is surrounded by a \mbox{3.2\,kton} water-Cherenkov outer detector~(OD) that serves as a tag for cosmic-ray muons.

The $\overline{\nu}_{e}$'s are detected by the inverse beta decay reaction, \mbox{$\overline{\nu}_{e}+p\rightarrow e^{+}+n$}.
The detector is sensitive to a small fraction of $\overline{\nu}_{e}$'s from the decay chains of $^{238}$U and $^{232}$Th above the threshold, and insensitive to $^{40}$K decays due to their lower energies.
The delayed coincidence of the correlated events from the prompt $e^{+}$ and the delayed neutron capture by a proton with the 2.2\,MeV $\gamma$-ray emission which takes place $\sim$200 $\mu$s later than the prompt signal is a powerful tool to reduce backgrounds.
The $\overline{\nu}_{e}$ event-selection is optimized to minimize background effects.
Its criteria, the detection efficiencies, and the systematic uncertainties are the same as in the previous study~\citep{Gando2011,Gando2013}.
We have been observing $\overline{\nu}_{e}$'s with KamLAND since 2002.
The KamLAND LS was purified in 2007 and 2008, and a $\beta\beta$-decay source was deployed from 2011 to 2015 to search for neutrinoless $\beta\beta$-decay (KamLAND-Zen 400)~\citep{Gando2016}.
The $\beta\beta$-decay source consists of 13 tons of Xe-loaded liquid scintillator (Xe-LS), contained in a 3.08-m-diameter inner balloon (IB) at the center of the detector.
In 2018, the KamLAND-Zen experiment was upgraded to KamLAND-Zen 800 by doubling the amount of xenon and installing a larger inner balloon of radius 1.9\,m~\citep{Gando2020,zencollaboration2021nylon}.

\begin{figure}
    \centering
    \includegraphics[width=1\linewidth]{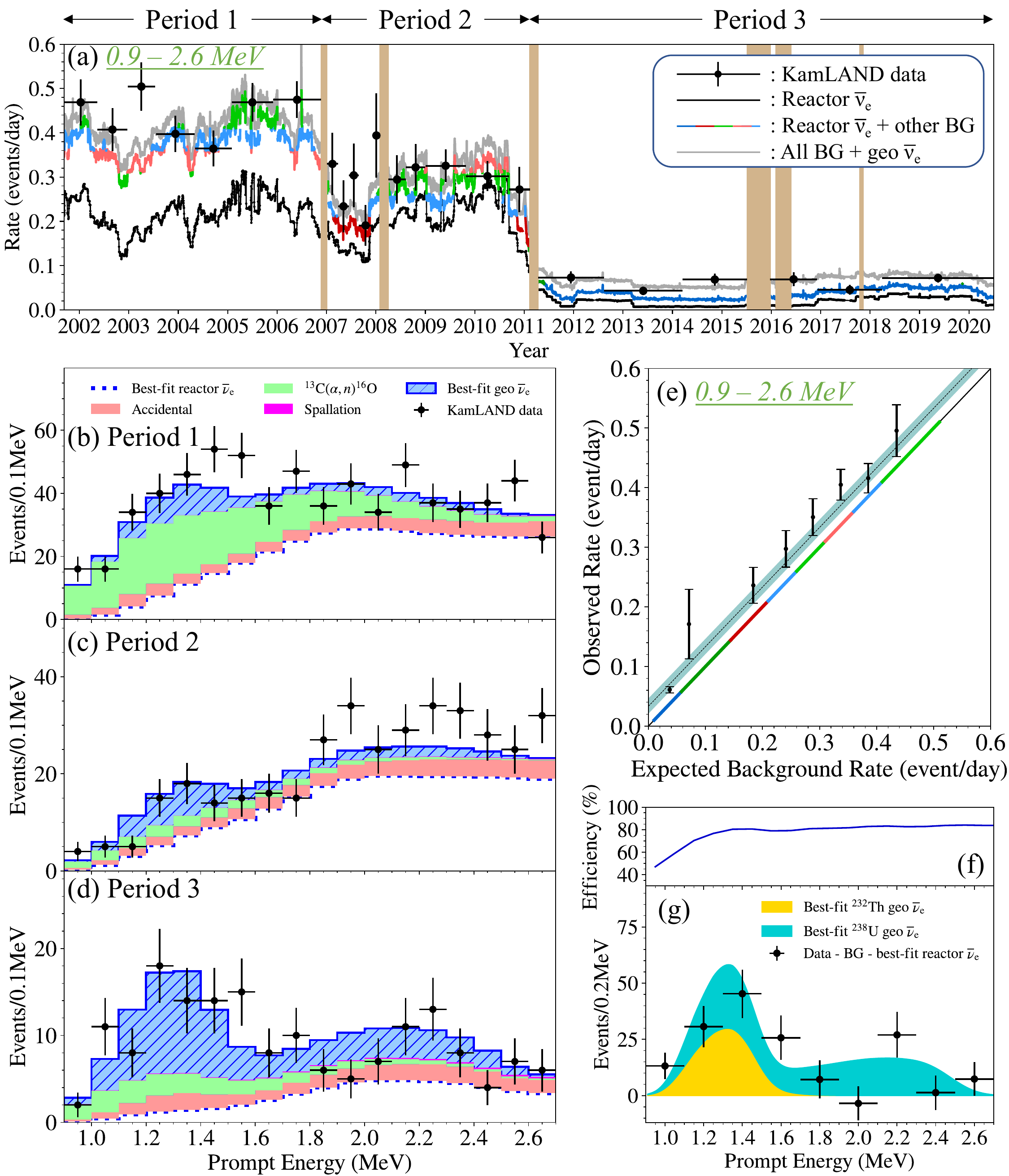}
    \caption{Observed event profiles at KamLAND with prompt energies between $0.9\,{\rm MeV}$ and $2.6\,{\rm MeV}$. (a)~Time variation of observed and expected event rates.
    The colored segments of the reactor $\overline{\nu}_\mathrm{e}$ + other BG are binned as the expected background rate.
    The vertical bands correspond to data periods that are not used in this analysis due to liquid scintillator purification activities in 2007 and 2008, KamLAND-Zen installation work in 2011, 2016, and 2018, and OD refurbishment work in 2016.
    (b,c,d)~Prompt energy spectrum for Period 1, 2, and 3.
    (e)~Comparison of observed rate versus expected background rate.
    The data are integrated within the each colored segments in (a). The best fit value of the geoneutrino contribution from the whole data-set analysis is shown as a black dashed line together with the $\pm$2$\sigma$ C.L. (shaded region).
    (f) Energy-dependent selection efficiency averaged over the whole data-set.
    (g) Background-subtracted observed energy spectrum and the best-fit $^{238}\mathrm{U}$ and $^{232}\mathrm{Th}$ geo $\overline{\nu}_\mathrm{e}$.}
    \label{fig:energy_time_spectrum}
\end{figure}

The data collected between March 9, 2002 and December 31, 2020, represents a total live-time of 5227\,days.
The data set is divided into three periods as shown in Figure~\ref{fig:energy_time_spectrum}.
\mbox{Period 1} (1486\,days live-time) refers to data recorded before a LS purification campaign that started in May 2007 and continued into 2009.
\mbox{Period 2} (1151\,days live-time) refers to data taken during and after the LS purification campaign, and \mbox{Period 3} (2590\,days live-time) denotes the data taken after installing the IB.
The number of target protons in the spherical fiducial volume of radius 6.0\,m is estimated to be $(5.98 \pm 0.13) \times 10^{31}$, resulting in a total exposure of $(6.39 \pm 0.14) \times 10^{32}$ target-proton-years.
Data taken during the LS purification activities exhibited increased PMT noise and were excluded from the data set.
During the $\beta\beta$-decay source deployment, i.e. from 2011 Aug. to 2015 Dec. and from 2018 May to present, the $\overline{\nu}_{e}$ analysis volume is restricted to a volume outside the IB to avoid backgrounds from the IB material.

The background due to reactor $\overline{\nu}_{e}$ has the largest impact on the geoneutrino measurement because of the shared energy range.
Since it is the by-product of the generation of electricity from nuclear reactors, it's flux is outside of the experiment's control.
We work with the power companies to calculate the reactor neutrino flux, spectrum and time variation from instantaneous thermal power and records of burnup and refueling for all commercial reactors in Japan.
We calculate the contributions from Korean and other regional reactors based on electric power records reported by the Korean power companies and IAEA~\citep{opex2020iaea}, assuming a 10\% error on the $\overline{\nu}_{e}$ flux.
A key input is the $\overline{\nu}_{e}$ spectral shape per isotope fission.
Recent improvements in the fission models introduces a $\sim$3\% upward shift~\citep{Mueller2011,Huber2011} relative to previous calculations~\citep{Schreckenbach1985,Hahn1989}, leading to a predicted $\overline{\nu}_{e}$ flux somewhat higher than that measured in previously reported short-baseline experiments operating within 1 km of reactors.
In addition, recent short-baseline experiments see an excess of events in the region of $4-6$\,MeV by $\sim$10\%, indicating the need to revise the reactor $\overline{\nu}_{e}$ modeling~\citep{PhysRevLett.123.111801}.
This is an active area of study; however, the origins of these discrepancies are still unknown.

In order to minimize systematic uncertainties from reactor modeling, we use the $\overline{\nu}_{e}$ spectrum constructed from the $e^{+}$ energy spectrum measured by the Daya Bay experiment to normalize the prediction~\citep{An2017}.
They provide a database containing a set of energy spectra in units of $\overline{\nu}_{e}$ per total fission, and the corresponding covariance matrix extracted from the unfolding method.
The relative fission yield depends on the reactor operation parameters, so we apply a small correction to the spectrum of each reactor,
\begin{eqnarray}
\label{equation:correction}
S_{reac.} = S_{DB} + \sum_{i} (\alpha_{i}^{reac.} - \alpha_{i}^{DB}) S_{i}, 
\end{eqnarray}
where $S_{DB}$ is the $\overline{\nu}_{e}$ spectrum in Daya Bay, $\alpha_{i}$ and $S_{i}$ are the fractional fission rate and the $\overline{\nu}_{e}$ spectrum of isotope $i$ from the models~\citep{Mueller2011,Huber2011}.
The second term for each reactor is typically small.
The contributions to $S_{DB}$ from the non-equilibrium effect and spent nuclear fuel for Daya Bay reactors, which are not appropriate for other reactors, are subtracted based on the estimate provided in~\citet{An2017}. 
Instead, we added the evaluated contributions from the long-lived, non-equilibrium fission products $^{90}$Sr, $^{106}$Ru, $^{144}$Ce, $^{97}$Zr, $^{132}$I, and $^{93}$Y~\citep{Kopeikin2001}, based on the history of fission rates for each isotope. The estimated contribution to the total $\overline{\nu}_{e}$ events is only $(0.7 \pm 0.3)\%$.

Taking into account the effect of neutrino oscillation, the expected number of reactor $\overline{\nu}_{e}$ events in the geoneutrino region (defined as $0.9~{\rm MeV} < E_{\rm p} < 2.6~{\rm MeV}$) is \mbox{$608^{+11}_{-13}$}.
The prompt event energy, $E_{\rm p}$, including the positron's kinetic energy and its annihilation energy, is related to the incident antineutrino energy, $E_{\overline{\nu}_{e}}$, by $E_{\rm p} = E_{\overline{\nu}_{e}} - 0.8$~MeV.
The relative contribution of long-lived products is $2.4 \pm 1.2\%$.
Non-neutrino backgrounds are dominated by radio-impurity in the detector components.
The $^{13}{\rm C}(\alpha,{\it\ n})^{16}{\rm O}$ reaction triggered by $\alpha$-decay of $^{210}$Po in the LS is a background, which was significantly reduced by two LS purification campaigns.
The $^{13}{\rm C}(\alpha,{\it\ n})^{16}{\rm O}$ background is estimated with satisfactory precision, as discussed in previous works~\citep{Araki2005b, Harissopulos2005, McKee2008, Abe2008, Gando2011, Gando2013}.
A recent publication reports newly measured $^{13}{\rm C}(\alpha,{\it\ n})^{16}{\rm O}$ cross-section for the 2nd excited state (6.13\,MeV $\gamma$)~\citep{Febbraro2020}; however, this data is not used because of clear disagreement with the event rate of $^{210}{\rm Po}^{13}{\rm C}$ source calibration data in KamLAND~\citep{McKee2008}.
Including minor contributions from accidental coincidences originating from radio-impurity in the LS, cosmic-ray-muon-induced radioactive isotopes, fast neutrons, and atmospheric neutrinos, the total number of background events in the geoneutrino region is estimated to be $923^{+35}_{-35}$.

After applying all the selection cuts, 1178 candidates remain in the geoneutrino region.
By interpreting the excess over the expected backgrounds as the geoneutrino signal, we obtain an estimate of \mbox{$255 \pm 49$} geoneutrinos without energy and time information.
We perform geoneutrino spectroscopy to estimate the uranium and thorium
individual contributions by means of an unbinned maximum-likelihood analysis based on the event rate, energy, and time information, and simultaneous fits of geoneutrinos and reactor $\overline{\nu}_{e}$'s including the effect of neutrino oscillation. 
Furthermore, the analysis employs the constraint on the oscillation parameters from solar~\citep{Cleveland_1998,Abdurashitov_2009,Bellini_2011,Hosaka_2006,Aharmim_2013}, accelerator~\citep{Abe2011,Adamson2011} and reactor~\citep{Abe_2012,An_2013,Ahn2012} neutrino data.

Figure~\ref{fig:energy_time_spectrum}(a) shows the event rate variation in the geoneutrino region; a constant contribution over the other time-varying components is interpreted as the geoneutrino signal.
The geoneutrino contribution clearly appears in the correlation plot of the expected background rate versus the observed rate (Figure~\ref{fig:energy_time_spectrum}(e)).
The ratios of the geoneutrino signal over background for the Periods 1, 2, and 3 are 0.10, 0.15, and 0.74, respectively, indicating the high geoneutrino detection sensitivity in Period 3.
The event time information is highly advantageous for discriminating the reactor $\overline{\nu}_{e}$ background which varies with time, as well as the accidental and $^{13}{\rm C}(\alpha,{\it n})^{16}{\rm O}$ backgrounds.

The event energy is used to discriminate backgrounds, and for determining the relative contributions of geoneutrinos from $^{238}$U and $^{232}$Th, whose spectral shapes are characterized by the effective end-point of 3.272\,MeV and 2.254\,MeV in neutrino energy, respectively. 
Figure~\ref{fig:energy_time_spectrum}(b-d) show the best-fit energy spectrum for each data-taking period.
The $^{13}{\rm C}(\alpha,{\it n})^{16}{\rm O}$ that were prominent in Period 1 (Figure~\ref{fig:energy_time_spectrum}(b)) was drastically reduced by the purification campaigns in Period 2 (Figure~\ref{fig:energy_time_spectrum}(c)).
As shown in Figure~\ref{fig:energy_time_spectrum}(d), the energy spectrum for Period 3 shows a clear peak at $\sim$1.3\,MeV consistent with the geoneutrino shape, and also shows the low background rate.

Figure~\ref{fig:energy_time_spectrum}(g) shows the livetime-weighted average of the background-subtracted observed energy spectrum overlaid with the best-fit geoneutrino spectrum accompanied with the energy dependent selection efficiency over the whole data-set (Figure~\ref{fig:energy_time_spectrum}(f)).
The best-fit geoneutrino signals are $117^{+41}_{-39}$ and $58^{+25}_{-24}$ events from $^{238}$U and $^{232}$Th, respectively. The null hypothesis of no $^{238}$U signals is disfavored at 3.3$\sigma$ confidence level (C.L.).
Based on the fit described above, we obtained $174^{+31}_{-29}$ geoneutrinos in total.
Fixing a Th/U mass ratio of 3.9 predicted by compositional analysis of chondrites~\citep{McDonough1995},
the total number of geoneutrinos is constrained to $183^{+29}_{-28}$. This corresponds to a $\overline{\nu}_{e}$ flux of \mbox{$3.4^{+0.5}_{-0.5} \times 10^{6} \, {\rm cm^{-2}s^{-1}}$} from $^{238}$U and $^{232}$Th at the Earth's surface.
The null hypothesis is disfavored at 8.5$\sigma$ C.L. from the $\Delta \chi^{2}$-profile (Figure~\ref{fig:combined_figure_FofU_FofTh_models}(b)).

The hypothesis that there may exist
a natural nuclear fission reactor in the Earth's interior, a so-called georeactor~\citep{Herndon2003}, can be tested using KamLAND data.
The energy spectra are fit by adding a constant flux from the hypothetical georeactor assuming a fission ratio of commercial power reactors with the averaged oscillation effect.
The geoneutrinos from U and Th are allowed to vary for the spectrum fitting.
The KamLAND data give a limit on the georeactor power of <1.26 TW at 90\% C.L.
The present result with the recent reactor-free data have improved the previous by KamLAND limit~\citep{Gando2013} by a factor of 2.5.

\section{Radiogenic Heat Estimation and Discussion of the Earth Composition Models}

The primary motivation of geoneutrino measurements is the estimation of radiogenic heat production in the Earth.
Our previous studies concluded that heat from radioactive decay contributes to about half of the Earth's total heat flow~\citep{Gando2011,Gando2013}, indicating global secular cooling under the assumed simple Earth model.
Thanks to the accumulation of the recent reactor-off data, we have gained a significant reduction in the geoneutrino flux uncertainties, permitting 
both better constraints on radioactive heat and also separation of heat producing elements.

The radiogenic heat can be estimated from the observed geoneutrino flux on the basis of the framework of the Earth model.
As a first approximation, we assume homogeneous distributions of U and Th in each region of the shell structure.
In the history of planetary differentiation, the highly incompatible elements, such as U, Th, and K, have been concentrated near the surface, and depleted in deeper regions. The mantle is the region of primary interest.
In the differentiation process of the crust and mantle, the repeated melting near the surface continuously extracts U and Th from the mantle material.
On the other hand, subduction returns those surface materials enriched in heat-producing elements to the mantle.
Hence the compositional heterogeneity of the mantle depends on the spatial range of the global mantle convection.
A representative mantle convection model for the Earth remains an open question, given observations and modeling studies ($\it{e.g.}$, seismic tomography~\citep{Fukao2013}, viscosity~\citep{Rudolph2015}, and structure~\citep{Ballmer2017}). 
In this work, we focus on the mantle with simple assumptions to constrain the model.

If we assume U and Th are absent in the core, the radiogenic heat from the mantle can be estimated from the observed geoneutrino flux ($\Phi$) after subtracting the crustal contribution in an Earth model.
Thus we obtain the total radiogenic heat from $^{238}$U and $^{232}$Th,
\begin{eqnarray}
\label{equation:mantleheat}
Q^{\rm U,Th} \, [{\rm TW}] & = & Q^{\rm U,Th}_{\rm crust} + Q^{\rm U,Th}_{\rm mantle} \\
 & = & Q^{\rm U,Th}_{\rm crust} + (\Phi^{\rm U,Th} \, [{\rm cm^{-2}s^{-1}}] - \Phi^{\rm U,Th}_{\rm crust} \, [{\rm cm^{-2}s^{-1}}]) \frac{\mathrm{d}Q^{\rm U,Th}_{\rm mantle}}{\mathrm{d}\Phi^{\rm U,Th}_{\rm mantle}}, 
\end{eqnarray}
where $r^{\rm U,Th} \equiv \mathrm{d}Q^{\rm U,Th}_{\rm mantle}/\mathrm{d}\Phi^{\rm U,Th}_{\rm mantle}$ is the ratio of the radiogenic heat to the geoneutrino flux determined by the spatial distributions of U and Th in the mantle.
For the crustal contribution, we use the geological reference model~\citep{Enomoto2007}, predicting that the radiogenic heat from the crust is $Q^{\rm U}_{\rm crust} = 3.4$\,TW and $Q^{\rm Th}_{\rm crust} = 3.6$\,TW. 
The uncertainties on the crustal geoneutrino flux from U and Th are assigned for the following calculations on the radiogenic heat.
Uncertainties on U and Th abundances in the crust are available for the upper and middle continental crusts in~\citep{Rudnick2014}, and for the lower continental crust in~\citep{Sramek2016}.
The total uncertainty of the crustal geoneutrino flux are estimated to be 24\% and 11\% for U and Th, respectively, conservatively assuming full correlations between each crustal layer.
In addition, the Th/U mass ratio in the continental crust, evaluated by the time-integrated Pb isotopic ratio (3.95$^{+0.19}_{-0.13}$)~\citep{Wipperfurth2018}, is used to provide constraints on U and Th abundance in the following radiogenic heat calculations.
Assuming the homogeneous distributions of U and Th in the mantle, the radiogenic heat is calculated to be $Q^{\rm U} = 3.3^{+3.2}_{-0.8}$\,TW and $Q^{\rm Th} =12.1^{+8.3}_{-8.6}$\,TW, where the 1$\sigma$ lower limits allow for unphysical solutions \mbox{($Q_{\rm mantle}<0$)}.
Combining the U and Th contributions with their anti-correlated errors, we obtain a radiogenic heat of $Q^{\rm U} + Q^{\rm Th} = 15.4^{+8.3}_{-7.9}$\,TW for the case with Th/U left free.  
If we fix the Th/U mass ratio at 3.9, the radiogenic heat estimates are improved to $Q^{\rm U} = 5.1^{+2.4}_{-2.0}$\,TW, $Q^{\rm Th} = 5.9^{+2.7}_{-2.2}$\,TW, and $Q^{\rm U} + Q^{\rm Th} = 10.6^{+5.2}_{-4.2}$\,TW, indicating some excess above the crustal contributions, {\it i.e.} a positive mantle contribution.
Using a geochemical study estimate of the $^{40}\mathrm{K}$ radiogenic heat to be $\sim 4$\,TW~\citep{Arevalo2009}, the Earth's total radiogenic heat is estimated to be $\sim14.6$\,TW using the reference model~\citep{Enomoto2007}.

If the reference model~\citep{Enomoto2007} is replaced with a modern model~\citep{Wipperfurth2020}, which predicts 
larger crustal contributions by 6\% and 9\% for $^{238}$U and $^{232}$Th, 
respectively, the radiogenic heat estimations decreased to $Q^{\rm U} = 4.3^{+4.3}_{-2.2}$\,TW, $Q^{\rm Th} = 4.8^{+4.8}_{-2.4}$\,TW, and $Q^{\rm U} + Q^{\rm Th} = 9.2^{+9.0}_{-4.6}$\,TW. 
Given such variation of crustal contribution in the reference models, the discussion on the Earth's radiogenic heat based on geoneutrino measurements can strongly depend on the accuracy of crust representation, especially in the area close around the detector (<500~km).
Our reference model~\citep{Enomoto2007} treats the local non-uniformity as the uncertainty to global modeling.
A recent study reported that a stochastic modeling method of 3-D compositional distribution in the crust was developed and applied to the Japan arc for a better understanding of the local crustal contribution~\citep{Takeuchi2019}.
Insights from geological perspectives provide the alternative descriptions of the near field lithospheres surrounding the detector, which is providing more accurate models for interpreting the geoneutrino flux measurement~\citep{Sammon2022}.

There are three groups of competing estimates for the BSE composition, Low-Q, Middle-Q, and High-Q models, categorized in~\citet{Sramek2013, Sramek2016}. 
These models have different predictions of the radiogenic heat in the Earth, {\it e.g.} Low-Q (10-15 TW), Middle-Q (17-22 TW), and High-Q models (> 25 TW).
The High-Q model~\citep{Turcotte2002} is motivated by Earth's heat budget leading to realistic mantle convection, requiring high radiogenic abundances.
The Middle-Q model~\citep{McDonough1995} uses compositional estimates based on the analyses of CI carbonaceous chondrites and terrestrial samples in consideration of elemental enrichment by planetary differentiation.
The Low-Q model~\citep{Javoy2010} assumes a mantle composition similar to that of enstatite chondrites, resulting in low radiogenic abundances.
In addition to the amount of the radiogenic heat, there has been controversy regarding the distribution of the U and Th in the mantle.
Two types of hypothesis are assumed with the aim of testing the models using the measured geoneutrino flux: the ``homogeneous hypothesis'' introducing U and Th uniformly throughout the mantle ($1/r^{\rm U + Th} = 1.14 \times 10^{5}$), and the ``sunken-layer hypothesis'' putting all of the U and Th at the mantle-core interface ($1/r^{\rm U + Th} = 8.5 \times 10^{4}$).
The measured geoneutrino flux at KamLAND is compared with the expectations from the BSE compositional models through radiogenic heat.

\begin{figure}
    \centering
    \vspace{-5mm}
    \includegraphics[width=0.75\columnwidth]{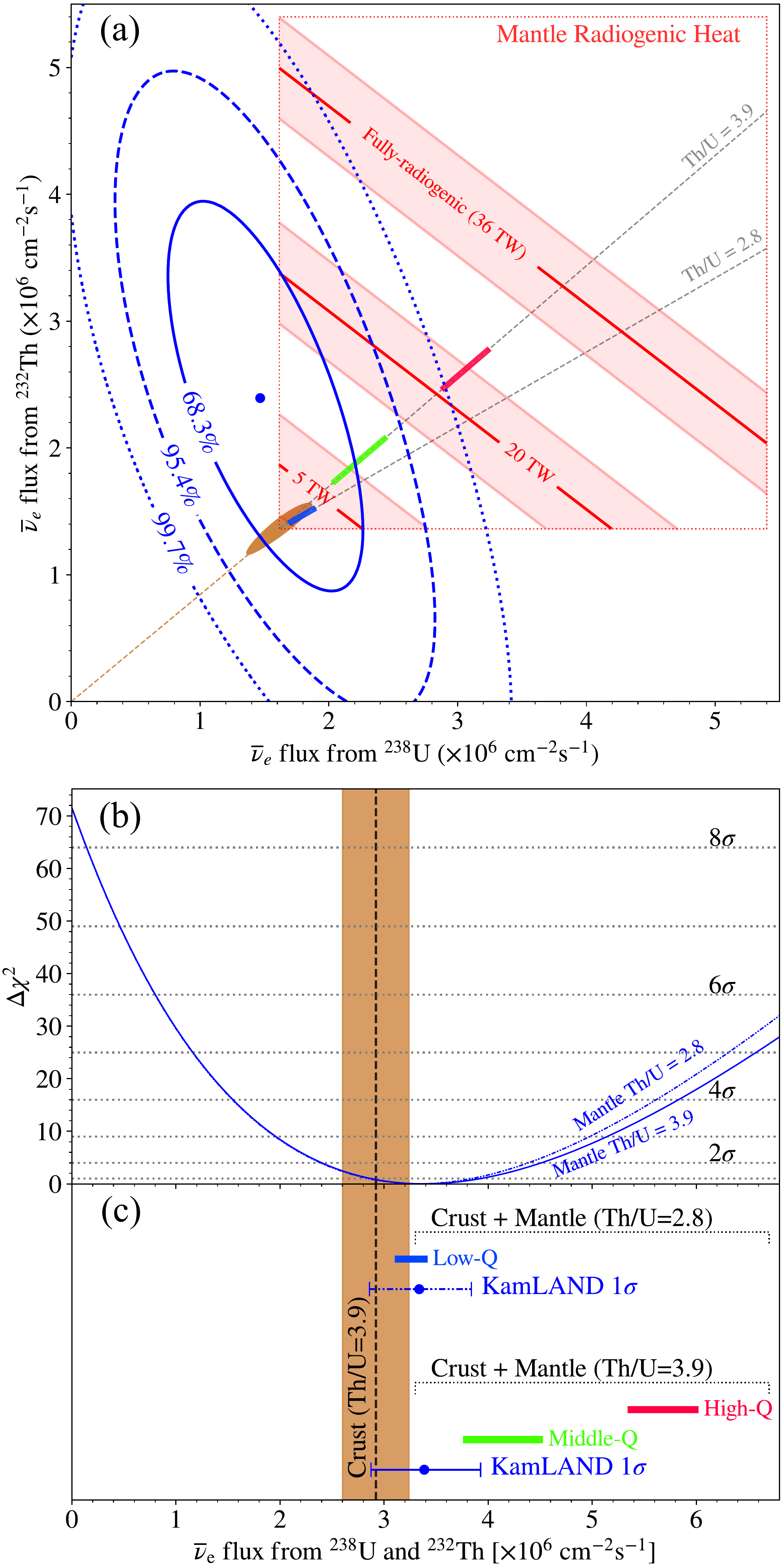}
    \caption{$\Delta\chi^2$-profiles of observed geoneutrino flux with mantle models. (a) C.L. contours and best fit point for $^{238}\mathrm{U}$ and $^{232}\mathrm{Th}$ geoneutrino flux from KamLAND's data (blue lines and point) with crustal prediction (brown ellipse), radiogenic heat from the mantle (three red lines and bands) and different mantle models (blue, green and red as depicted in (c)). (b) Projection on the gray dashed lines rising to the right in (a) of mantle mass ratio Th/U=3.9 for Middle-Q and High-Q models (blue line) and Th/U=2.8 for Low-Q model (blue dash-dotted line) starting projection from the center of the crustal contribution. (c) Comparison of the measured data and the mantle models on each mantle mass ratio.}
    \label{fig:combined_figure_FofU_FofTh_models}
\end{figure}

Figure~\ref{fig:combined_figure_FofU_FofTh_models}(a) shows the $\Delta\chi^2$-contours for the observed geoneutrino flux from $^{238}$U and $^{232}$Th along with the Mantle Radiogenic Heat panel.
The red lines downward to the right on the Mantle Radiogenic Heat panel correspond to the selected mantle ragiogenic heat from $^{238}$U and $^{232}$Th with the homogeneous hypothesis, and the accompanying red band indicates the crustal uncertainty.
The fully radogenic model value on the Mantle Radiogenic Heat panel is obtained by subtracting the crust and $^{40}$K contributions from the surface heat flow.
The color-filled bars represent the expected mantle radiogenic heat from the different mantle models, High-Q (red), Middle-Q (green) and Low-Q (blue), with the uncertainties of heat-producing element abundances \citep{Sramek2013} re-interpreted with our reference crust model~\citep{Enomoto2006}.
They are on the gray dashed lines rising to the right which represent the expected Th/U mass ratio in the mantle of 3.9 (High-Q and Middle-Q models) and 2.8 (Low-Q model) \citep{Sramek2013}.
The brown ellipse indicates the crustal uncertainties which are estimated using the uncertainties in $^{238}$U and $^{232}$Th concentrations \citep{Rudnick2014, Sramek2016} with the constraint on the crustal Th/U mass ratio estimated by the time-integrated Pb isotopic ratio (3.95$^{+0.19}_{-0.13}$) \citep{Wipperfurth2018}.
The center of the brown ellipse is equal to the predicted crustal geoneutrino contribution of our reference model~\citep{Enomoto2006}.
It is connected to the brown dashed line which describes Th/U ratio (3.9) in the crust estimated by the reference model \citep{Enomoto2007}

Furthermore, Figure~\ref{fig:combined_figure_FofU_FofTh_models}(b) shows the $\Delta\chi^2$-profiles of geoneutrino flux projected to the gray lines in Figure~\ref{fig:combined_figure_FofU_FofTh_models}(a) corresponding to Mantle Th/U=2.8 (for Low-Q model) and Th/U=3.9 (for Middle-Q and High-Q models).
Figure~\ref{fig:combined_figure_FofU_FofTh_models}(c) illustrates the comparison of the measured data and the different mantle models.
The KamLAND data is consistent with the expectations from the chondrite-based (Low-Q and Middle-Q) models, but indicates  some incompatibility with the High-Q model.
Assuming Gaussian errors for the crustal contribution and for the BSE abundances on the gray tilted line to the right of the Th/U mass ratio at 3.9, we find that the High-Q model's prediction is disfavored at 99.76\% C.L. with the homogeneous hypothesis, and at 97.9\% C.L. with the sunken-layer hypothesis.

The incompatibility between the High-Q model and the chondrite-based (Low-Q and Middle-Q) models favored by the KamLAND data may indicate the need to modify the conventionally parameterized convection model, as discussed in~\citet{Korenaga2006}.
In such thermal evolution models, the fraction of the global heat production from radioactive decays, the so-called ``Urey ratio'', is an important parameter.
The mantle contribution alone is referred to as the ``convective Urey ratio'' \citep{Korenaga2008}.
Assuming extra mantle heat contributions of 3.0\,TW from other heat producing elements ($^{40}$K and $^{235}$U)~\citep{Arevalo2009, Enomoto2006}, we find that the convective Urey ratio is $0.13^{+0.15}_{-0.06}$ with the homogeneous hypothesis.
The fully radiogenic model (Urey ratio of 1) is disfavored at $5.2\sigma$ with the homogeneous hypothesis, and at $4.0\sigma$ even with the sunken-layer hypothesis, so the indication that there exists a substantial contribution from the Earth's primordial heat supply in~\citet{Gando2011} is strengthened.
The present-day contributions from individual heat producing elements determine the past radiogenic heat through the history of the Earth, depending on their lifetimes.
In future experiments, geoneutrino spectroscopy including $^{40}$K will provide a very important missing piece of information for geochemical and geophysical studies.

\section{Summary}

Based on the updated KamLAND data including the recent reactor-off period, we have employed geoneutrino spectroscopy to measure the U and Th geoneutrino contributions.
We have demonstrated the capability to spectroscopically identify heat producing elements, and provide the best constraint to date on the radioactive heat contribution to the bulk Earth.
The determination of the mantle radiogenic heat is a matter of primary interest for geoscience.
For the case of existing detectors in crustal locations, the experimental uncertainty on the mantle geoneutrino flux is mainly caused by the statistical subtraction of the crustal contribution, depending on the geological model.
In the future, if a detector in an oceanic location with small crustal contribution is realized, multi-site flux data will better determine the individual geoneutrino contributions from the crust and mantle.
Although $^{40}$K is one of the primary heat producing elements, albeit minor, in the thermal history of the Earth, existing detectors are insensitive to $^{40}$K.
New detection techniques need to be pursued to tighten neutrino constraints on the total geoneutrino heat production.

\section*{Data Availability Statement}
Our numerical data on this paper's figures are available in a repository at Tohoku University  "https://www.awa.tohoku.ac.jp/KamLAND/geonu2022/data-release.html".  These include the estimated energy spectra for backgrounds and $\overline{\nu}_\mathrm{e}$ candidate events, and $\Delta\chi^{2}$ map of geoneutrino flux from $^{238}$U and $^{232}$Th.

\acknowledgments
The \mbox{KamLAND} experiment is supported by JSPS KAKENHI Grant Numbers 21000001, 26104002, and 19H05803; the Dutch Research Council (NWO); and under the U.S. Department of Energy (DOE) Grant No.\,DE-AC02-05CH11231, National Science Foundation (NSF) Grant No.\,2021964, 
as well as other DOE and NSF grants to individual institutions.
The reactor data are provided by courtesy of the following electric associations in Japan: Hokkaido, Tohoku, Tokyo, Hokuriku, Chubu, Kansai, Chugoku, Shikoku, and Kyushu Electric Power Companies, Japan Atomic Power Company, and Japan Atomic Energy Agency. 
The Kamioka Mining and Smelting Company has provided service for activities in the mine. We acknowledge the support of NII for SINET.

\bibliography{GeoNeutrino}

\begin{thebibliography}{50}
\providecommand{\natexlab}[1]{#1}
\expandafter\ifx\csname urlstyle\endcsname\relax
  \providecommand{\doi}[1]{doi:\discretionary{}{}{}#1}\else
  \providecommand{\doi}{doi:\discretionary{}{}{}\begingroup
  \urlstyle{rm}\Url}\fi

\bibitem[{\textit{Abdurashitov et~al.}(2009)}]{Abdurashitov_2009}
Abdurashitov, J.~N., et~al. (2009), Measurement of the solar neutrino capture
  rate with gallium metal. iii. results for the 2002--2007 data-taking period,
  \textit{Phys. Rev. C}, \textit{80}, 015,807,
  \doi{10.1103/PhysRevC.80.015807}.

\bibitem[{\textit{Abe et~al.}(2011)}]{Abe2011}
Abe, K., et~al. (2011), Indication of electron neutrino appearance from an
  accelerator-produced off-axis muon neutrino beam, \textit{Phys. Rev. Lett.},
  \textit{107}, 041,801, \doi{10.1103/PhysRevLett.107.041801}.

\bibitem[{\textit{Abe et~al.}(2008)}]{Abe2008}
Abe, S., et~al. (2008), Precision measurement of neutrino oscillation
  parameters with {KamLAND}, \textit{Phys. Rev. Lett.}, \textit{100}(22),
  221803.

\bibitem[{\textit{Abe et~al.}(2012)}]{Abe_2012}
Abe, Y., et~al. (2012), {Reactor ${\overline{\ensuremath{\nu}}}_{e}$
  disappearance in the Double Chooz experiment}, \textit{Phys. Rev. D},
  \textit{86}, 052,008, \doi{10.1103/PhysRevD.86.052008}.

\bibitem[{\textit{Adamson et~al.}(2011)}]{Adamson2011}
Adamson, P., et~al. (2011), Improved search for muon-neutrino to
  electron-neutrino oscillations in {MINOS}, \textit{Phys. Rev. Lett.},
  \textit{107}, 181,802, \doi{10.1103/PhysRevLett.107.181802}.

\bibitem[{\textit{Adey et~al.}(2019)}]{PhysRevLett.123.111801}
Adey, D., et~al. (2019), Extraction of the $^{235}\mathrm{U}$ and
  $^{239}\mathrm{Pu}$ antineutrino spectra at {Daya Bay}, \textit{Phys. Rev.
  Lett.}, \textit{123}, 111,801, \doi{10.1103/PhysRevLett.123.111801}.

\bibitem[{\textit{Aharmim et~al.}(2013)}]{Aharmim_2013}
Aharmim, B., et~al. (2013), Combined analysis of all three phases of solar
  neutrino data from the {Sudbury Neutrino Observatory}, \textit{Phys. Rev. C},
  \textit{88}, 025,501, \doi{10.1103/PhysRevC.88.025501}.

\bibitem[{\textit{Ahn et~al.}(2012)}]{Ahn2012}
Ahn, J.~K., et~al. (2012), Observation of reactor electron antineutrinos
  disappearance in the {RENO} experiment, \textit{Phys. Rev. Lett.},
  \textit{108}, 191,802, \doi{10.1103/PhysRevLett.108.191802}.

\bibitem[{\textit{An et~al.}(2013)}]{An_2013}
An, F.~P., et~al. (2013), Improved measurement of electron antineutrino
  disappearance at {Daya Bay}, \textit{Chin. Phys. C}, \textit{37}(1), 011,001,
  \doi{10.1088/1674-1137/37/1/011001}.

\bibitem[{\textit{An et~al.}(2017)}]{An2017}
An, F.~P., et~al. (2017), Improved measurement of the reactor antineutrino flux
  and spectrum at {Daya Bay}, \textit{Chin. Phys. C}, \textit{41}(1), 013,002.

\bibitem[{\textit{Araki et~al.}(2005)}]{Araki2005b}
Araki, T., et~al. (2005), Experimental investigation of geologically produced
  antineutrinos with {KamLAND}, \textit{Nature}, \textit{436}, 499--503.

\bibitem[{\textit{Arevalo~Jr. et~al.}(2009)\textit{Arevalo~Jr., McDonough, and
  Luong}}]{Arevalo2009}
Arevalo~Jr., R., W.~F. McDonough, and M.~Luong (2009), The {K/U} ratio of the
  silicate {Earth}: {Insights} into mantle composition, structure and thermal
  evolution, \textit{Earth and Planet. Sci. Lett.}, \textit{278}(3-4),
  361--369.

\bibitem[{\textit{Ballmer et~al.}(2017)}]{Ballmer2017}
Ballmer, M.~D., et~al. (2017), Persistence of strong silica-enriched domains in
  the earth's lower mantle, \textit{Nature Geosci.}, \textit{10}, 236--240,
  \doi{doi.org/10.1038/ngeo2898}.

\bibitem[{\textit{Bellini et~al.}(2010)}]{Bellini2010}
Bellini, G., et~al. (2010), Observation of geo-neutrinos, \textit{Phys. Lett.
  B}, \textit{687}(4-5), 299--304.

\bibitem[{\textit{Bellini et~al.}(2011)}]{Bellini_2011}
Bellini, G., et~al. (2011), Precision measurement of the $^{7}\mathrm{Be}$
  solar neutrino interaction rate in {Borexino}, \textit{Phys. Rev. Lett.},
  \textit{107}, 141,302, \doi{10.1103/PhysRevLett.107.141302}.

\bibitem[{\textit{Cleveland et~al.}(1998)}]{Cleveland_1998}
Cleveland, B.~T., et~al. (1998), Measurement of the solar electron neutrino
  flux with the {Homestake} chlorine detector, \textit{The Astrophysical
  Journal}, \textit{496}(1), 505--526, \doi{10.1086/305343}.

\bibitem[{\textit{Davies and Davies}(2010)}]{Davies2010}
Davies, J.~H., and D.~R. Davies (2010), Earth's surface heat flux,
  \textit{Solid Earth}, \textit{1}(1), 5--24.

\bibitem[{\textit{Enomoto}(2006)}]{Enomoto2006}
Enomoto, S. (2006), Experimental study of geoneutrinos with {KamLAND},
  \textit{Earth, Moon, and Planets}, \textit{99}(1), 131--146.

\bibitem[{\textit{Enomoto et~al.}(2007)\textit{Enomoto, Ohtani, Inoue, and
  Suzuki}}]{Enomoto2007}
Enomoto, S., E.~Ohtani, K.~Inoue, and A.~Suzuki (2007), Neutrino geophysics
  with {KamLAND} and future prospects, \textit{Earth and Planet. Sci. Lett.},
  \textit{258}, 147--159.

\bibitem[{\textit{Febbraro et~al.}(2020)}]{Febbraro2020}
Febbraro, M., et~al. (2020), New $^{13}\mathrm{C}(\alpha, n)^{16}\mathrm{O}$
  cross section with implications for neutrino mixing and geoneutrino
  measurements, \textit{Phys. Rev. Lett.}, \textit{125}(6), 062,501--,
  \doi{10.1103/PhysRevLett.125.062501}.

\bibitem[{\textit{Fukao and Obayashi}(2013)}]{Fukao2013}
Fukao, F., and M.~Obayashi (2013), Subducted slabs stagnant above, penetrating
  through, and trapped below the 660 km discontinuity, \textit{J. Geophys.
  Res.}, \textit{118}, 5920--5938, \doi{doi.org/10.1016/j.pepi.2021.106815}.

\bibitem[{\textit{Gando et~al.}(2011)}]{Gando2011}
Gando, A., et~al. (2011), Partial radiogenic heat model for earth revealed by
  geoneutrino measurements, \textit{Nature Geosci}, \textit{4}(9), 647--651.

\bibitem[{\textit{Gando et~al.}(2013)}]{Gando2013}
Gando, A., et~al. (2013), Reactor on-off antineutrino measurement with
  {KamLAND}, \textit{Phys. Rev. D}, \textit{88}(3), 033,001.

\bibitem[{\textit{Gando et~al.}(2016)}]{Gando2016}
Gando, A., et~al. (2016), Search for majorana neutrinos near the inverted mass
  hierarchy region with {KamLAND-Zen}, \textit{Phys. Rev. Lett.},
  \textit{117}(8), 082,503.

\bibitem[{\textit{Gando}(2020)}]{Gando2020}
Gando, Y. (2020), {First results of KamLAND-Zen 800}, \textit{Journal of
  Physics: Conference Series}, \textit{1468}(1), 012,142,
  \doi{10.1088/1742-6596/1468/1/012142}.

\bibitem[{\textit{Gando et~al.}(2021)}]{zencollaboration2021nylon}
Gando, Y., et~al. (2021), The nylon balloon for xenon loaded liquid
  scintillator in {KamLAND-Zen} 800 neutrinoless double-beta decay search
  experiment, \textit{JInst}, \textit{16}(08), P08,023,
  \doi{10.1088/1748-0221/16/08/p08023}.

\bibitem[{\textit{Hahn et~al.}(1989)}]{Hahn1989}
Hahn, A.~A., et~al. (1989), Antineutrino spectra from {$^{241}{\rm Pu}$ and
  $^{239}{\rm Pu}$} thermal neutron fission products, \textit{Phys. Lett. B},
  \textit{218}(3), 365--368.

\bibitem[{\textit{Harissopulos et~al.}(2005)}]{Harissopulos2005}
Harissopulos, S., et~al. (2005), Cross section of the {$^{13}{\rm
  C}(\alpha,n)^{16}{\rm O}$} reaction: {A} background for the measurement of
  geo-neutrinos, \textit{Phys. Rev. C}, \textit{72}(6), 062,801,
  \doi{10.1103/PhysRevC.72.062801}.

\bibitem[{\textit{Herndon}(2003)}]{Herndon2003}
Herndon, J.~M. (2003), Nuclear georeactor origin of oceanic basalt {$^{3}{\rm
  He}$/$^{4}{\rm He}$}, evidence, and implications, \textit{Proc. of Natl.
  Acad. Sci. U.S.A.}, \textit{100}(6), 3047--3050.

\bibitem[{\textit{Hosaka et~al.}(2006)}]{Hosaka_2006}
Hosaka, J., et~al. (2006), Solar neutrino measurements in {Super-Kamiokande-I},
  \textit{Phys. Rev. D}, \textit{73}, 112,001,
  \doi{10.1103/PhysRevD.73.112001}.

\bibitem[{\textit{Huber}(2011)}]{Huber2011}
Huber, P. (2011), Determination of antineutrino spectra from nuclear reactors,
  \textit{Phys. Rev. C}, \textit{84}(2), 024,617.

\bibitem[{\textit{IAEA}(2021)}]{opex2020iaea}
IAEA (2021), {Official Website of IAEA, International Atomic Energy Agency,
  Operating Experience with Nuclear Power Stations in Member States},
  https://www.iaea.org.

\bibitem[{\textit{Javoy et~al.}(2010)}]{Javoy2010}
Javoy, M., et~al. (2010), The chemical composition of the earth: Enstatite
  chondrite models, \textit{Earth and Planet. Sci. Lett.}, \textit{293}(3--4),
  259--268.

\bibitem[{\textit{Kopeikin et~al.}(2001)\textit{Kopeikin, Mikaelyan, and
  Sinev}}]{Kopeikin2001}
Kopeikin, V.~I., L.~A. Mikaelyan, and V.~V. Sinev (2001), Inverse beta decay in
  a nonequilibrium antineutrino flux from a nuclear reactor, \textit{Phys. of
  At. Nucl.}, \textit{64}(5), 849--854.

\bibitem[{\textit{Korenaga}(2006)}]{Korenaga2006}
Korenaga, J. (2006), Archean geodynamics and the thermal evolution of earth,
  \textit{Archean Geodynamics and Environments}, \textit{164}, 7--32.

\bibitem[{\textit{Korenaga}(2008)}]{Korenaga2008}
Korenaga, J. (2008), Urey ratio and the structure and evolution of {Earth's}
  mantle, \textit{Rev. Geophys.}, \textit{46}(2), RG2007.

\bibitem[{\textit{Lyubetskaya and Korenaga}(2007)}]{Lyubetskaya2007}
Lyubetskaya, T., and J.~Korenaga (2007), Chemical composition of {Earth's}
  primitive mantle and its variance: 2. {Implications} for global geodynamics,
  \textit{J. Geophys. Res.}, \textit{112}(B3), B03,212.

\bibitem[{\textit{McDonough and Sun}(1995)}]{McDonough1995}
McDonough, W.~F., and S.~Sun (1995), The composition of the {Earth},
  \textit{Chem. Geol.}, \textit{120}, 223--253.

\bibitem[{\textit{McKee et~al.}(2008)\textit{McKee, Busenitz, and
  Ostrovskiy}}]{McKee2008}
McKee, D.~W., J.~K. Busenitz, and I.~Ostrovskiy (2008), A {$^{13}{\rm
  C}(\alpha,n)^{16}{\rm O}$} calibration source for {KamLAND}, \textit{Nucl.
  Instrum. and Meth. A}, \textit{587}(2-3), 272--276.

\bibitem[{\textit{Mueller et~al.}(2011)}]{Mueller2011}
Mueller, T.~A., et~al. (2011), Improved predictions of reactor antineutrino
  spectra, \textit{Phys. Rev. C}, \textit{83}(5), 054,615.

\bibitem[{\textit{Rudnick and Gao}(2014)}]{Rudnick2014}
Rudnick, R.~L., and S.~Gao (2014), \textit{Treatise on Geochemistry:
  Composition of the Continental Crust {Vol.} 3}, pp. 1--51, Pergamon Press,
  Oxford.

\bibitem[{\textit{Rudolph et~al.}(2015)\textit{Rudolph, Leki{\'c}, and
  Lithgow-Bertelloni}}]{Rudolph2015}
Rudolph, M.~L., V.~Leki{\'c}, and C.~Lithgow-Bertelloni (2015), Viscosity jump
  in earth's mid-mantle, \textit{Science}, \textit{350}(6266), 1349--1352,
  \doi{10.1126/science.aad1929}.

\bibitem[{\textit{Sammon and McDonough}(2022)}]{Sammon2022}
Sammon, L.~G., and W.~F. McDonough (2022), Quantifying earth's radiogenic heat
  budget, \textit{Earth and Planetary Science Letters}, \textit{593}, 117,684,
  \doi{https://doi.org/10.1016/j.epsl.2022.117684}.

\bibitem[{\textit{Schreckenbach et~al.}(1985)\textit{Schreckenbach, Colvin,
  Gelletly, and Von~Feilitzsch}}]{Schreckenbach1985}
Schreckenbach, K., G.~Colvin, W.~Gelletly, and F.~Von~Feilitzsch (1985),
  Determination of the antineutrino spectrum from {$^{235}{\rm U}$} thermal
  neutron fission products up to 9.5 {MeV}, \textit{Phys. Lett. B},
  \textit{160}(4-5), 325--330.

\bibitem[{\textit{{\v S}r{\'a}mek et~al.}(2013)\textit{{\v S}r{\'a}mek,
  McDonough, Kite, Leki{\'c}, Dye, and Zhong}}]{Sramek2013}
{\v S}r{\'a}mek, O., W.~F. McDonough, E.~S. Kite, V.~Leki{\'c}, S.~T. Dye, and
  S.~Zhong (2013), Geophysical and geochemical constraints on geoneutrino
  fluxes from earth's mantle, \textit{Earth and Planet. Sci. Lett.},
  \textit{361}(0), 356--366.

\bibitem[{\textit{{\v S}r{\'a}mek et~al.}(2016)\textit{{\v S}r{\'a}mek,
  Roskovec, Wipperfurth, Xi, and McDonough}}]{Sramek2016}
{\v S}r{\'a}mek, O., B.~Roskovec, S.~A. Wipperfurth, Y.~Xi, and W.~F. McDonough
  (2016), Revealing the earth's mantle from the tallest mountains using the
  {Jinping} neutrino experiment, \textit{Scientific Reports}, \textit{6},
  33,034, \doi{doi.org/10.1038/srep33034}.

\bibitem[{\textit{Takeuchi et~al.}(2019)}]{Takeuchi2019}
Takeuchi, N., et~al. (2019), Stochastic modeling of 3-d compositional
  distribution in the crust with {Bayesian} inference and application to
  geoneutrino observation in {Japan}, \textit{Physics of the Earth and
  Planetary Interiors}, \textit{288}, 37--57,
  \doi{https://doi.org/10.1016/j.pepi.2019.01.002}.

\bibitem[{\textit{Turcotte and Schubert}(2002)}]{Turcotte2002}
Turcotte, D.~L., and G.~Schubert (2002), \textit{Geodynamics, Applications of
  Continuum Physics to Geological Problems, second ed., Cambridge University
  Press}.

\bibitem[{\textit{Wipperfurth et~al.}(2020)\textit{Wipperfurth, {\v
  S}r{\'a}mek, and McDonough}}]{Wipperfurth2020}
Wipperfurth, S.~A., O.~{\v S}r{\'a}mek, and W.~F. McDonough (2020), Reference
  models for lithospheric geoneutrino signal, \textit{Journal of Geophysical
  Research}, \textit{125}, e2019JB018,433, \doi{doi.org/10.1029/2019JB018433}.

\bibitem[{\textit{Wipperfurth et~al.}(2018)}]{Wipperfurth2018}
Wipperfurth, S.~A., et~al. (2018), Earth's chondritic {Th/U}: Negligible
  fractionation during accretion, core formation, and crust--mantle
  differentiation, \textit{Earth and Planetary Science Letters}, \textit{498},
  196--202, \doi{https://doi.org/10.1016/j.epsl.2018.06.029}.

\end{thebibliography}

\listofchanges

\end{document}